\documentclass[prd,showpacs]{revtex4}

\usepackage{graphicx}

\usepackage{bm}
\newcommand{\vc}[1]{\bm{#1}}

\newcommand{\vf}[1]{\overrightarrow{#1}}

\newcommand{\ieps}[1]{\includegraphics[height=3in]{#1}}

\begin{document}

\title{Optimal generalization of power filters for gravitational wave bursts, from single to multiple detectors} 
\author{Julien Sylvestre}
\affiliation{LIGO Laboratory, California Institute of Technology,\\MS 18-34, Pasadena, CA 91125, USA.}

\email{jsylvest@ligo.caltech.edu}

\date{\today}

\begin{abstract}
Searches for gravitational wave signals which do not have a precise model describing the shape of their waveforms are often performed using power detectors based on a quadratic form of the data.
A new, optimal method of generalizing these power detectors so that they operate coherently over a network of interferometers is presented.
Such a mode of operation is useful in obtaining better detection efficiencies, and better estimates of the position of the source of the gravitational wave signal.
Numerical simulations based on a realistic, computationally efficient hierarchical implementation of the method are used to characterize its efficiency, for detection and for position estimation.
The method is shown to be more efficient at detecting signals than an incoherent approach based on coincidences between lists of events.
It is also shown to be capable of locating the position of the source. 
\end{abstract}


\maketitle

\section{Introduction} \label{intro}
Six kilometer-scale laser interferometers designed to observe gravitational waves (GW) with unprecedented sensitivities should complete or approach the end of their commissioning in the year 2003.
Three are operated in North America by the LIGO Laboratory \cite{LIGO}, two in Europe by the Virgo \cite{Virgo} and the GEO600 \cite{GEO} projects, on one in Asia by the TAMA300 \cite{TAMA} project.
A collaborative analysis of the data collected by these instruments provides the best prospects for detecting and analyzing GW events of astronomical origin.

The focus of this article will be on ``bursts'' of gravitational radiation, loosely defined as GW of duration of the order of a few seconds at most, and present in a frequency range overlapping at least partially with the bandwidth of the interferometers (10 Hz $\alt f \alt$ 1 kHz).
Other types of signals that will not be discussed here include continuous GW from rotating neutron stars, and a stochastic background of GW of cosmological origin.

Arguments based on the astrophysics and on the dynamics of the sources of GW bursts show that the detection of these signals will be challenging, as the signals will be deeply buried in the instrumental noise \cite{Thorne300yrs}.
Consequently, a significant research effort is currently on-going to develop and study efficient algorithms for the detection and the characterization of the elusive GW signals.
An important fraction of the literature on the subject concerns signals with a precisely known form \cite{matched}.
The knowledge of the signal allows the construction of a phase coherent filter (the Wiener or matched filter) which is known to be optimal for signal detection.
Only the coalescence of compact binaries and possibly the ringdown of excited black holes should be detectable using matched filtering.
For the particular case of compact binaries, it is known that a coherent analysis using data from all the interferometers of the international network will improve the detection prospects noticeably \cite{Pai,Finn}, although the computational cost of such an analysis might be prohibitive \cite{Pai2}.
In addition, it was shown in \cite{Sylvestre} that the use of the Advanced LIGO detectors and of the Virgo interferometer cooperatively might allow the localization of the GW source with enough accuracy to permit its observation with electromagnetic instruments, thus complementing with information about the thermodynamics of the source the information on its dynamics provided by the GW.

The remainder of this article will be concerned only with GW signals that are not known with enough precision to allow matched filtering.
The algorithms that have been proposed in the literature to detect these signals fall into two general categories: time-domain filters, and power detectors.
Time-domain filters \cite{timeDomain} rely on the development of a small bank of linear filters which are expected to cover relatively well the space of possible GW signals.
They offer the advantages of speed, simplicity, and possibly ease of interpretation, but might lack the robustness and efficacy of power detectors.
Only the latter will be discussed here.
The power detectors threshold on some non-linear measure of the data, often constructed from a time-frequency representation of the signal \cite{tfplane,tfplane1,epower,vicere,tfclusters,wnoise}.
They have been shown to be optimal for the detection of signals with especially poor waveform descriptions.

All power detectors were explicitly designed and implemented to process data from the different interferometers of the world-wide network independently.
Under this mode of operation, it is expected that event lists are generated individually from the data stream provided by each interferometer, and are later compared to form coincidences based on temporal, frequency, or more general information.
This {\it incoherent} approach should not yield the maximum efficacy, in part because GW bursts in individual interferometers have to be rather loud to register with the power detector and to give accurate estimates of their start time, duration, frequency band, amplitude, etc., all of which might be needed by the coincidence gate.
The alternative is to combine all data streams first, and then run a burst detector on the {\it synthetic} data stream so produced.
I implement this {\it coherent} approach as a generalization of the power detectors developed to date for single interferometers, by calculating the optimal way to combine any number of interferometer data streams into a single time-series, such that when this time-series is fed to a single interferometer power detector, a larger signal-to-noise ratio is obtained than for any other combination of the data streams.
This brings to the already implemented and well-characterized power detectors the benefits of a network coherent analysis, which include improved sensitivity, and the ability to precisely locate the source position on the sky.

\subsection{Summary of Results}
The coherent power filter ({\tt CPF}) algorithm presented in this paper involves the following steps:
\begin{enumerate}
\item A point on a grid in parameter space is chosen, where the parameters are the source angular position, and two numbers describing the plus and cross polarizations of the GW signal. One of these numbers is the ratio of the power in the cross polarization to the power in the plus polarization, and the other is the amount of overlap between the two polarizations, approximately measured by integrating the product of the two waveforms over time.
\item Given the source position and the network topology, the data streams from all interferometers are time-shifted to align the GW signals to a common origin in time.
\item Every data stream is multiplied by a scale factor, and all the data streams are added together to form a synthetic time-series. The choice of the scale factors depends on the network topology and on the four parameters chosen in Step 1.
\item The synthetic time-series is processed by a power detector, and the power measurement is recorded. 
\item If all points of the grid in parameter space have been visited, the algorithm exits. If the maximum of the power measurements exceeds a certain threshold, a detection is announced, and the parameters that gave the largest power measurement are returned as an estimate of the source parameters. 
\item Back to Step 1.
\end{enumerate}
A more detailed discussion of this algorithm is presented in section \ref{Algorithm}.

The scale factors in Step 3 are chosen so that the value of the signal-to-noise ratio (SNR) is maximized.
It is quite significant that only two parameters in addition to the source position are required to perform this maximization.
Geometrically, this can be understood by realizing that the signals in all interferometers, after being properly time-shifted, are linear combinations of the two polarization waveforms, and therefore lie in a hyperplane spanned by these two polarization waveforms.
Consequently, a knowledge of the ratio of the lengths of the two polarization waveforms and of their angle with respect to each other, together with a knowledge of the beam-patterns of all interferometers, is sufficient to determine the signals in all interferometers, up to an overall scale factor, and up to the orientation of the hyperplane.
However, neither of these two pieces of information are needed to calculate how to linearly combine the signals to get maximum power in the synthetic time-series.

The validity of the geometrical picture and the actual conclusion that only four parameters are required to perform the SNR maximization depends critically on the right choice of the position dependent time-shifts.
As shown in section \ref{Algorithm}, in the specific case where the cross-correlation function of the plus and the cross polarizations does not have an extremum at zero time lag, there are no formal guarantees that the {\tt CPF} algorithm will converge to the right source parameters.
Physically, this results from possible interactions (or cross-terms) between the plus and the cross polarizations which cannot be properly handled by the coherent algorithm.
This does not affect the detection performances of {\tt CPF}, but in some cases is significant for source position estimations.
As discussed in section \ref{Algorithm}, a number of canonical sources do not satisfy this condition exactly, so that a careful study of the position systematic errors is needed.
For the difficult case where the two polarizations are long monochromatic signals with a phase difference of a quarter of a cycle, it is shown in section \ref{systematics} that a correction for this systematic error can be implemented such that for $\sim 25\%$ of the sky the systematic error is negligible, while that for about $50\%$ of the sky it is too large to allow any position estimation.
This correction procedure only requires the additional knowledge of a quantity which is closely related to the characteristic frequency of the signal.

The performances of the {\tt CPF} algorithm are explored empirically in section \ref{simulations} through numerical simulations.
All the experiments are limited to the three interferometer network (the HLV network) consisting of the LIGO interferometer near Hanford, Washington, the LIGO interferometer near Livingston, Louisiana, and the Virgo interferometer in Italy.
The signal is short (1/16 s) and narrow band (25 Hz), and is assumed to originate from a position along the northern hemisphere normal to the HLV plane.
All experiments are performed with the {\tt TFCLUSTERS} \cite{tfclusters} algorithm as the single interferometer power detector.
A realistic, computationally efficient hierarchical implementation of the {\tt CPF} algorithm is shown to offer better detection performances than a incoherent approach which uses only coincidences between events generated by independent {\tt TFCLUSTERS} operating on the three interferometers.
It is also shown that the {\tt CPF} algorithm can be used to estimate the position of the source of GW.
When the GW signal has four times more power in its plus polarization than in its cross polarization, roughly one quarter of all trials lead to a position estimate that is within one degree from the true source position, for signals with reasonable amplitudes.
The ability to pinpoint the source location is debilitated by the misalignment of Virgo with respect to the LIGO detectors, the reduction of the signal-to-noise ratio, and the reduction of the ratio of the power in the plus and in the cross polarizations.

\section{Notation}

Let bold characters denote time-series; whether these time-series are continuous in time or discretely sampled will be immaterial in the following discussion.
It is assumed that a GW is observed with a network of $N$ independent detectors.
Calibrated data corresponding to measurements of the GW strain in all detectors are denoted $\vc{y}_i$, $i=1,2, ... , N$.
The noises $\vc{n}_i$ are assumed to be additive, so
\begin{equation}
\vc{y}_i = F^+_i(\theta, \phi, \psi) T[\Delta_i(\theta, \phi)] \vc{s}_+ + F^\times_i(\theta, \phi, \psi) T[\Delta_i(\theta, \phi)] \vc{s}_\times + \vc{n}_i, \label{eq:sigmodel}
\end{equation}
where $\vc{s}_+$ and $\vc{s}_\times$ are the two polarizations of the GW signal, $F^+_i$, $F^\times_i$ are the beam-pattern functions\cite{Thorne300yrs} of the $i^{\rm th}$ detector, and $T(\Delta)$ denotes the time-shift operator; for time-series with continuous time, for instance, $T(\Delta)\vc{x}(t) = \vc{x}(t - \Delta)$.
The beam-pattern functions depend on the two angles describing the source position (the right ascension and the declination, denoted $\theta$ and $\phi$ respectively), and on the polarization angle $\psi$.
The time-shift at the $i^{\rm th}$ detector, denoted $\Delta_i$, is the same for the two polarizations, and depends only on the source position on the sky.
The frame in which $\vc{s}_+$ and $\vc{s}_\times$ are defined is irrelevant since the waveforms are not assumed to be known a priori; a rotation of that frame is equivalent to a change in $\vc{s}_+$, $\vc{s}_\times$ and $\psi$.
The parameters $\theta, \phi, \psi, \vc{s}_+$, and $\vc{s}_\times$, and the derived quantities $F^+_i(\theta, \phi, \psi)$, $F^\times_i(\theta, \phi, \psi)$, and $\Delta_i(\theta, \phi)$, will be used below to describe the parameters of a real source which is assumed to be present in the data, and which we are trying to detect.

The scalar product between two time-series is denoted $\vc{x}\cdot\vc{y}$.
For time-series with continuous time, it is defined as
\begin{equation}
\vc{x}\cdot\vc{y} = \int_{-\infty}^\infty\int_{-\infty}^\infty \vc{x}(t_x) Q(t_x, t_y)  \vc{y}(t_y) dt_x dt_y, \label{eq:dotproduct}
\end{equation}
and similarly for time-series with discrete time.
The kernel $Q$ can be viewed as a filter applied to the time-series in order to detect more efficiently a particular signal or to modify the character of the noise, for instance.
The square of the norm of a time-series, also called its power, is denoted $|\vc{x}|^2$, and is defined by $|\vc{x}|^2 = \vc{x} \cdot \vc{x}$.

The noise in each of the $N$ interferometers is only assumed to be wide-sense stationary \cite{WSS}, i.e., it does not have to be Gaussian or white.
The noises can always be made zero mean and independent by linear filtering \cite{KL}, so $E[\vc{y}_i \cdot T(\Delta)\vc{y}_j] = R_i(\Delta)$ if $i=j$ and is zero otherwise, for $R_i(\Delta)$ the autocorrelation of the noise, and $E[\cdot]$ denoting the expectation value of its argument.
As usual, the Fourier transform of the noise autocorrelation function is the noise power spectral density.

\section{Algorithm} \label{Algorithm}

The {\it synthetic response} of the network is denoted $\vc{Y}$, and is a simple linear combination of the time-shifted individual detector responses:
\begin{equation}
\vc{Y} = \sum_{i=1}^N a_i T(\delta_i) \vc{y}_i, \label{eq:syntheticresponse}
\end{equation}
for some set of real coefficients $a_i$ and time shifts $\delta_i$, $i=1, ..., N$, which are arranged in two vectors, $\vf{a}$ and $\vf{\delta}$, respectively.
Note that the $\delta_i$ are the trial time delays used in the data analysis, and the algorithm defined below is used to estimate these delays so that they are close from the real delays $\Delta_i(\theta, \phi)$ corresponding to a source located at position $(\theta, \phi)$.
The {\it network power} $\hat{P}$ is the estimate of the power in the GW signal, according to our norm definition, i.e., $\hat{P} = |\vc{Y}|^2$.

The motivation behind this particular design is that it is the simplest generalization of the numerous power detectors for single interferometers described in the literature, which in many cases have already been implemented and characterized. 
In practical terms, a software code can be designed to compute the synthetic response for a network of interferometers, and these data can be fed to a power detector, as if they were data from a single interferometer, in order to measure the network power.
The kernel of the dot product used for the computation of the network power is then determined by the single interferometer power detector used to process the synthetic response.
Different power detectors are efficient for detecting different types of signals, so this generality of the synthetic response approach is very economical in terms of code development.
Some single interferometer power detectors provide a non-linear measure of the power; this is not a serious limitation given the algorithm structure defined above, for the power measurements are all very nearly linear for detectable signals.

Let $\delta_{ij} = \delta_i - \delta_j + \Delta_i - \Delta_j$ denotes the error on the estimated time-of-flight between detectors $i$ and $j$.
The network power can be expanded as
\begin{equation}
\hat{P} = \zeta + \eta,
\end{equation}
where the signal term is given by
\begin{eqnarray}
\zeta = \sum_{i,j=1}^N a_i a_j [ F^+_i F^+_j R_{++}(\delta_{ij}) + F^+_i F^\times_j R_{+\times}(\delta_{ij}) + \nonumber \\ 
F^\times_i F^+_j R_{\times+}(\delta_{ij}) + F^\times_i F^\times_j R_{\times\times}(\delta_{ij})], \label{eq:zeta}
\end{eqnarray}
and where the noise term is given by
\begin{equation}
\eta = \sum_{i,j=1}^N a_i a_j [ T(\delta_i + \Delta_i)(F^+_i \vc{s}_+ + F^\times_i \vc{s}_\times) \cdot T(\delta_j + \Delta_j) \vc{n}_j + T(\delta_i + \Delta_i) \vc{n}_i \cdot T(\delta_j + \Delta_j) \vc{n}_j ].
\end{equation}
The signal correlation functions are given by
\begin{equation}
R_{ij}(t_i - t_j) = T(t_i) \vc{s}_i \cdot T(t_j) \vc{s}_j,
\end{equation}
for $i,j = +$ or $\times$.

The signal-to-noise ratio $\rho$ is defined as
\begin{equation}
\rho^2 = \frac{\zeta}{E[\eta]},
\end{equation}
where the expectation of the noise can be rewritten as
\begin{equation}
E[\eta] = \sum_{i = 1}^N a_i^2 \sigma^2_i,
\end{equation}
where the noise variance is $\sigma^2_i = R_i(0)$.
For fixed noises and signals, the signal-to-noise ratio depends only on $\vf{a}$ and $\vf{\delta}$.
It can therefore be maximized for a given choice of the source parameters ($\theta, \phi, \psi, \vc{s}_+, \vc{s}_\times$) by varying $\vf{a}$ and $\vf{\delta}$.
Let $\vf{a}_m(\theta', \phi', \psi', \vc{s}_+', \vc{s}_\times')$ and $\vf{\delta}_m(\theta', \phi', \psi', \vc{s}_+', \vc{s}_\times')$ denote those values of $\vf{a}$ and $\vf{\delta}$ which maximize $\rho^2$ for some source parameters identified by primes to differentiate them from the true source parameters.
The following algorithm is then defined:
\begin{enumerate}
\item Pick a set of trial source parameters $\theta', \phi', \psi', \vc{s}_+', \vc{s}_\times'$.
\item Compute $\vf{a}_m(\theta', \phi', \psi', \vc{s}_+', \vc{s}_\times')$ and $\vf{\delta}_m(\theta', \phi', \psi', \vc{s}_+', \vc{s}_\times')$.
\item Form the synthetic response $\vc{Y}$ from $\vf{a}_m$ and $\vf{\delta}_m$.
\item Estimate $\hat{P}(\vc{Y})$ using a single detector algorithm.
\item Retain the source parameters $\theta', \phi', \psi', \vc{s}_+', \vc{s}_\times'$ if $[\hat{P}(\vc{Y}) - E[\eta]]/E[\eta]$ is the largest to date.
\item Go back to Step 1.
\end{enumerate}
The expectation of $[\hat{P}(\vc{Y}) - E[\eta]]/E[\eta]$ is just $\rho^2$, so on average this algorithm will converge to the true parameters of the source.

The maximization problem for $\rho^2$ can be recast as the maximization of $\zeta$ subjected to the constraint that $E[\eta]$ is constant.
The normal equations are
\begin{eqnarray}
\sum_{j=1, j\neq i}^N a_j [ F^+_i F^+_j R_{++}'(\delta_{ij}) +  F^+_i F^\times_j R_{+\times}'(\delta_{ij}) + \nonumber \\
F^\times_i F^+_j R_{\times+}'(\delta_{ij}) + F^\times_i F^\times_j R_{\times\times}'(\delta_{ij}) - \nonumber \\
F^+_i F^+_j R_{++}'(\delta_{ji}) -  F^\times_i F^+_j R_{+\times}'(\delta_{ji}) - \nonumber \\
F^+_i F^\times_j R_{\times+}'(\delta_{ji}) - F^\times_i F^\times_j R_{\times\times}'(\delta_{ji}) ] = 0 \label{eq:normdelta}\\
\lambda a_i \sigma_i^2 + a_i [F^+_i F^+_i \vc{s}_+ \cdot \vc{s}_+ + (F^+_i F^\times_i + F^\times_i F^+_i) \vc{s}_+ \cdot \vc{s}_\times + F^\times_i F^\times_i \vc{s}_\times \cdot \vc{s}_\times] + \nonumber \\
\sum_{j=1, j\neq i}^N a_j [ F^+_i F^+_j R_{++}(\delta_{ij}) +  F^+_i F^\times_j R_{+\times}(\delta_{ij}) + \nonumber \\
F^\times_i F^+_j R_{\times+}(\delta_{ij}) + F^\times_i F^\times_j R_{\times\times}(\delta_{ij}) ] = 0, \label{eq:normalpha} 
\end{eqnarray}
where $\lambda$ is the Lagrange parameter for the constraint, $R_{ij}'(x) = dR_{ij}(x)/dx$, and $i = 1, 2, 3$.

Eq. (\ref{eq:normdelta}) can be simplified using the identity $R_{ij}(x) = R_{ji}(-x)$:
\begin{equation}
\sum_{j=1, j\neq i}^N a_j [ F^+_i F^+_j R_{++}'(\delta_{ij}) +  F^+_i F^\times_j R_{+\times}'(\delta_{ij}) + \nonumber \\
F^\times_i F^+_j R_{\times+}'(\delta_{ij}) + F^\times_i F^\times_j R_{\times\times}'(\delta_{ij})] = 0. \label{eq:normdeltaI} 
\end{equation}
If $R_{+\times}(x)$ has an extremum at $x=0$, i.e. if $R_{+\times}'(x)|_{x=0}=0$, then a solution to Eq. (\ref{eq:normdeltaI}) is $\delta_i = -\Delta_i$ for $i=1,...,N$, since $R_{++}(x)$ and $R_{\times\times}(x)$ are maximal at $x=0$.
Back into Eq. (\ref{eq:normalpha}), this solution gives
\begin{eqnarray}
\lambda a_i \sigma_i^2 + \sum_{j=1}^N a_j [F^+_i F^+_j \vc{s}_+ \cdot \vc{s}_+ + (F^+_i F^\times_j + F^\times_i F^+_j) \vc{s}_+ \cdot \vc{s}_\times + F^\times_i F^\times_j \vc{s}_\times \cdot \vc{s}_\times] = 0. \label{eq:normalphaII}
\end{eqnarray}
The choice of $\vf{a}$ to maximize the signal-to-noise ratio then depends on the angles $\theta, \phi, \psi$ as before, but now only on the two numbers $|\vc{s}_+| / |\vc{s}_\times|$ and $\vc{s}_+ \cdot \vc{s}_\times / |\vc{s}_+||\vc{s}_\times|$ instead of the full waveforms for the two polarizations.
These numbers are denoted $\Lambda_{+/\times}$ and  $\Lambda_{+\cdot\times}$, respectively.
Eq. (\ref{eq:normalphaII}) is an eigenvalue problem (with the Lagrange parameter $\lambda$ playing the role of the eigenvalue, and the weight vector $\vf{a}$ that of the eigenvector), and is straightforward to solve numerically.
The matrix of the eigenvalue problem is hermitian, so all eigenvalues are real; the eigenvector which maximizes Eq. (\ref{eq:zeta}) with $\delta_{ij}=0$ is picked to form the synthetic response.

To recapitulate, if $R_{+\times}$ has an extremum at zero, a synthetic waveform can be constructed from the data of all interferometers such that this waveform is optimal for its processing by power detectors, and the maximization of the detection statistic for parameter estimation can be performed over only four parameters: $\theta, \phi, \Lambda_{+/\times}$ and $\Lambda_{+\cdot\times}$.
I do not include $\psi$ in this list of parameters because it is completely degenerated with $\Lambda_{+/\times}$ and $\Lambda_{+\cdot\times}$.
In particular, changing the basis where the two polarizations are defined by a rotation (i.e., a redefinition of the polarization angle $\psi$) does not change the value of $R_{+\times}'(x)|_{x=0}$.
For the reminder of this paper, I will take $\psi \equiv 0$; this fixes the definition of $\vc{s}_+$ and $\vc{s}_\times$ with respect to the frame of reference of the detectors network.
The {\tt CPF} algorithm is then defined as followed:
\begin{enumerate}
\item Pick a set of trial source parameters $\theta'$, $\phi'$, $\Lambda'_{+/\times}$, and $\Lambda'_{+\cdot\times}$.
\item Compute $\vf{a}$ from Eq.(\ref{eq:normalphaII}), and set $\vf{\delta}=0$.
\item Form the synthetic response $\vc{Y}$ using Eq.(\ref{eq:syntheticresponse}).
\item Use a single interferometer power detector to calculate $\hat{P} = |\vc{Y}|^2$.
\item Retain the source parameters $\theta'$, $\phi'$, $\Lambda'_{+/\times}$, and $\Lambda'_{+\cdot\times}$ if $\hat{P}$ is the largest to date.
\item Go back to Step 1.
\end{enumerate}

If the signal is linearly polarized, it can be written as $\vc{s}_+ = \vc{s} \cos 2\psi$ and $\vc{s}_\times = -\vc{s} \sin 2\psi$, for some polarization angle $\psi$ and some waveform $\vc{s}$.
The eigenmatrix in Eq. (\ref{eq:normalphaII}) is then of rank one, and consequently only has a single non-trivial eigenvalue.
The corresponding eigenvector is given by $a_i = [F^+_i(\theta, \phi, 0) \cos 2\psi - F^\times_i(\theta, \phi, 0) \sin 2\psi] / \sigma_i^2 = F^+_i(\theta, \phi, \psi) / \sigma_i^2$ for $i=1,2,3$.
Hence, when the GW signal is linearly polarized, the signals from the different interferometers are optimally combined by weighting them with the ratio of their noise variance to the value of the beam pattern functions, weighted appropriately by the polarization angle.
The signals from the different interferometers of the network are therefore emphasized linearly in the observed power of the GW signal, and inversely in the power of the noise.

\begin{figure}
\begin{center}
\ieps{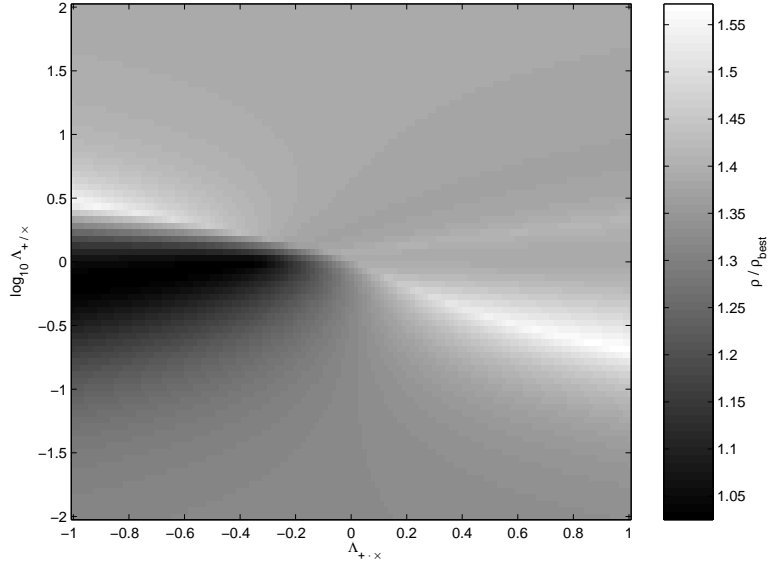}
\end{center}
\caption{The ratio $\rho / \rho_{\rm best}$ as a function of $\Lambda_{+/\times}$ and $\Lambda_{+\cdot\times}$, for a source along the northern hemisphere normal of the HLV plane.}
\label{fig:snrratios}
\end{figure}

Another interesting subcase involves {\em directed} searches: in that case, the position $(\theta,\phi)$ of a potential GW source is known precisely ($\delta_{ij} = 0$), and the goal is to be maximally sensitive to gravitational radiation from that source.
For $\hat{a}_i = a_i \sigma_i$, the maximization of the signal-to-noise ratio can be rewritten as the maximization of the quadratic form
\begin{equation}
\rho^2 = \sum_{i,j = 1}^N \hat{a}_i \hat{a}_j M_{ij}
\end{equation}
subjected to 
\begin{equation}
\sum_{i=1}^N \hat{a}_i^2 = 1,
\end{equation}
where the matrix $M$ has elements $M_{ij}$ given by
\begin{equation}
M_{ij} = \frac{|\vc{s}_+| |\vc{s}_\times|}{\sigma_i \sigma_j}\left[F^+_i F^+_j \Lambda_{+/\times} + (F^+_i F^\times_j + F^\times_i F^+_j) \Lambda_{+\cdot\times} + \frac{F^\times_i F^\times_j}{\Lambda_{+/\times}} \right].
\end{equation}
It is well known from Rayleigh's Principle \cite{noble} that the maximum value of $\rho^2$ is given by the largest eigenvalue of the matrix $M$.
This maximum signal-to-noise ratio can be compared to the signal-to-noise ratio that can be obtained using the best interferometer in the network, which is typical of the signal-to-noise ratio of an incoherent search.
It is given by
\begin{equation}
\rho^2_{\rm best} = \max_{i=1...N} M_{ii}.
\end{equation}
Considering the HLV network simplified so that all interferometers have the same noise level ($\sigma_i^2 = $constant), and fixing the beam-pattern functions by selecting a source along the northern hemisphere normal of the HLV plane, the ratio $\rho / \rho_{\rm best}$ varies between 1.03 and 1.57, depending on the value of $\Lambda_{+/\times}$ and $\Lambda_{+\cdot\times}$.
Figure \ref{fig:snrratios} shows that variation of the ratio $\rho / \rho_{\rm best}$ with $\Lambda_{+/\times}$ and $\Lambda_{+\cdot\times}$: when the signal has a large degree of linear polarization, the improvement is fairly large, with $\rho / \rho_{\rm best} \sim 1.4$.
When both polarizations roughly contain the same power, the improvement can be large ($\agt 1.5$) or very small ($\sim 1$), depending on the structure of the signal (i.e., on $\Lambda_{+\cdot\times}$).
If we add a detector at TAMA's location to the HLV network, with noise similar to the noise of the other interferometers, the ratio $\rho / \rho_{\rm best}$ varies between 1.15 and 1.79 for a source at the same location as before.
The four interferometer network works better than the three interferometer HLV network, although the improvement is somewhat limited by the unavoidable misalignment between instruments located on different continents.

If $R_{+\times}$ does not have an extremum at zero, the normal equations are not necessarily satisfied at $\delta_i = -\Delta_i$.
They become non-linear and rather complex to solve, but, more importantly, a full knowledge of the waveforms for the two polarizations is then necessary to obtain a solution.
Physically, this is a result of the fact that the two polarizations may interfere together constructively when shifted by a non-zero lag; this lag gets added to the estimated time delay between the interferometers in the network, and a systematic error in the source position estimate appears.
It is important to realize that this systematic error does not significantly reduce the detection capabilities of {\tt CPF}, but only its positioning ability.
Essentially, the convergence to a secondary maximum provides an alternative way to cross the detection threshold when the global maximum of the signal power is plagued by noise, such that the measured power at its position is not a global maximum of the measured network power.
It is nevertheless interesting to estimate the size of the position systematic error when the {\tt CPF} algorithm is used on a signal for which $R_{+\times}'(x)|_{x=0} = 0$ does not hold, i.e. when Eq. (\ref{eq:normalphaII}) is imposed as a solution.
This is the subject of the following subsection.

Before this question is addressed, however, it is worth examining the physical meaning of the condition $R_{+\times}'(x)|_{x=0}=0$.
As it was pointed out by \cite[Eqns. 2.30d, 4.3, and 5.18a]{ThorneTh}, the GW strain for slow-motion, weak gravity sources radiating mostly due to variations of their mass moments (by contrast to current moments) can be written as
\begin{equation}
h^{TT}(t) \propto \sum_{l=2}^\infty \sum_{m=-l}^l (\nabla \nabla Y^{lm})^{\rm STT} \frac{d^l}{dt^l} \int \rho(t-r/c) Y^{lm*} r^{2+l} dr d\Omega, \label{eq:hsph}
\end{equation}
where STT means ``symmetric transverse-traceless'', $Y^{lm}$ are the spherical harmonics, and $\rho$ is the mass density of the source.
A number of possible GW sources consist in anisotropic mass distributions that are rapidly rotating about a well-defined axis, and consequently radiate principally with $l=2$, $m=2$ (e.g., binaries, bar or fragmentation instabilities, longest live mode of a perturbed Kerr black hole, as noted by \cite{KM2002}).
The imaginary part of $Y^{22}$ is rotated by $\pi/4$ about the polar axis (the rotation axis) with respect to its imaginary part.
This results in the mass distribution being sampled at any given time according to two spatial patterns rotated by $\pi/4$ with respect to each other.
Since the GW are emitted at twice the rotation frequency of the source, this rotation angle produces a $\pi/2$ phase difference between the real and imaginary parts of the time dependent integral in Eq. (\ref{eq:hsph}).
As noted by \cite{KM2002}, the dependence of the polarization on the inclination angle comes from the pure-spin tensor harmonics $(\nabla \nabla Y^{lm})^{\rm STT}$, which for $l=2$ can be found in \cite{Mathews}.
For $m=2$,
\begin{equation}
(\nabla \nabla Y^{22})^{\rm STT} \propto (1 + \cos^2 \iota) e_+ + 2i \cos \iota \; e_\times,
\end{equation}
where $\iota$ is the inclination angle of the rotation axis with respect to the line of sight ($\iota = 0$ along the polar axis), and where $e_+$ and $e_\times$ are the unit linear-polarization tensors for the GW.
Consequently, the plus and cross polarization waveforms can be written as
\begin{eqnarray}
h_+ \propto (1 + \cos^2 \iota) \cos \Phi(t) \label{eq:pluspolarizations} \\
h_\times \propto 2 \cos \iota \sin \Phi(t), \label{eq:crosspolarizations}
\end{eqnarray}
where $\Phi(t)$ is some phase function.

By definition, equations (\ref{eq:pluspolarizations}) and (\ref{eq:crosspolarizations}) give
\begin{equation}
R_{+\times}(\tau) \propto \cos \iota (1 + \cos^2 \iota) \int_{-\infty}^\infty \int_{-\infty}^\infty dt_1 dt_2 Q(t_1, t_2) \cos \Phi(t_1) \sin \Phi(t_2 - \tau),
\end{equation}
so that
\begin{equation}
R_{+\times}'(x)\left|_{x=0}\right. \propto - \cos \iota (1 + \cos^2 \iota)  \int_{-\infty}^\infty \int_{-\infty}^\infty dt_1 dt_2 Q(t_1, t_2) \Phi'(t_2) \cos \Phi(t_1) \cos \Phi(t_2).
\end{equation}
In general, the phase $\Phi(t)$ will have at least a linear component [$\Phi(t) \sim \omega t$], so that
\begin{equation}
R_{+\times}'(x)\left|_{x=0}\right. \propto - \cos \iota (1 + \cos^2 \iota) \omega.
\end{equation}
As the system's polar axis becomes aligned with the line-of-sight, the assumption that $R_{+\times}'(x)|_{x=0}=0$ becomes progressively worst, and significant systematic position errors might appear, and will have to be accounted for, as discussed in the next section.
Some signals might also satisfy the condition $R_{+\times}'(x)|_{x=0}=0$. 
This is the case, for instance, if $\Phi(t)$ is an even function of $t$.

Sources which radiate predominantly in a $l=2$, $m=1$ mode will show a similar correlation between their plus and cross polarizations. 
In that case, the real and imaginary parts of $Y^{21}$ differ by a $\pi/2$ rotation about the polar axis, but have a $m=1$ symmetry, so that the waves are radiated at the spin frequency, and consequently the phase shift between the plus and cross polarizations is again $\pi/2$.
It might be that the dominating population of sources to be observed will not be dominated by rotation about a principal axis; as a result, the correlation between the plus and cross polarizations might be quite arbitrary.
In an axisymmetric core collapse, for instance, the $l=2$, $m=0$ mode dominates \cite{Muller}.
The angular response is $(\nabla \nabla Y^{20})^{\rm STT} \propto \sin^2 \iota \; e_+$, so the waves are linearly polarized.

\subsection{Systematic Position Errors} \label{systematics}
Eq. (\ref{eq:normdeltaI}) can be used to check the error on $\theta$ and $\phi$ by solving it for $\vf{\delta}$, with $\vf{a}$ obtained from the {\tt CPF} algorithm, i.e., from the solution of Eq. (\ref{eq:normalphaII}).
For every trial choice of $(\theta', \phi', \Lambda'_{+/\times}, \Lambda'_{+\cdot\times})$, the algorithm returns a set of weights $\vf{a}$ (by definition of {\tt CPF}, $\vf{\delta} = 0$).
However, the maximum of the signal-to-noise ratio occurs when the normal equations are satisfied; assuming that we let $\Lambda_{+/\times}$ and $\Lambda_{+\cdot\times}$ be varied freely in the maximization of $\hat{P}$, the values of $\theta$ and $\phi$ returned by this maximization will be those that solve Eq. (\ref{eq:normdeltaI}).

Consider the following parameterization of the signal cross-correlation functions:
\begin{eqnarray}
R'_{++}(t) = R'_{\times\times}(t) = -\omega_1^2 t + O(t^3) \\
R'_{+\times}(t) = \omega_0 - \omega_2^3 t^2 + O(t^4) \\
R'_{\times+}(t) = -\omega_0 + \omega_2^3 t^2 + O(t^4),
\end{eqnarray}
for some parameters $\omega_0$, $\omega_1$, and $\omega_2$.
If $\vc{s}_+$ and $\vc{s}_\times$ were long, nearly monochromatic signals of angular frequency $\omega$, with the same amplitude but a phase difference of $\pi/2$, for instance, the parameters could be chosen to be $\omega_0 = \omega_1 = \omega_2 \equiv \omega$.
In order to get an analytical solution, I linearize Eq. (\ref{eq:normdeltaI}) to obtain the first order equation
\begin{equation}
\sum_{j=1, j\neq i}^N a_j [ F_{ij} \omega_1^2 \delta^{(1)}_{ij} - \omega_0 G_{ij} ] = 0,
\end{equation}
where $\delta^{(1)}_{ij}$ represent first order errors between the true time delays and the delays returned by the algorithm, and where $F_{ij} = F^+_i F^\times_j + F^\times_i F^+_j$ and $G_{ij} = F^+_i F^\times_j - F^\times_i F^+_j$.
The solutions to this linear system of equations are degenerate; for $N=3$, they are
\begin{equation}
\frac{\omega_1^2}{\omega_0} \delta^{(1)}_{13} = -\frac{a_1 F_{12} G_{13} + a_2 F_{12} G_{23} + a_2 F_{23} G_{12} + a_3 F_{23} G_{13}}{a_1 F_{12} F_{13} + a_2 F_{12} F_{23} + a_3 F_{13} F_{23}} \label{eq:delta13}
\end{equation}
and
\begin{equation}
\frac{\omega_1^2}{\omega_0} \delta^{(1)}_{23} = -\frac{a_1 F_{12} G_{13} + a_2 F_{12} G_{23} - a_1 F_{13} G_{12} + a_3 F_{13} G_{23}}{a_1 F_{12} F_{13} + a_2 F_{12} F_{23} + a_3 F_{13} F_{23}} \label{eq:delta23}.
\end{equation}

The times-of-flight between two pairs of detectors are sufficient to triangulate the position of the source on the sky and to obtain its position $(\theta, \phi)$, up to a reflection with respect to the plane containing the three detectors.
The magnitude $l^{(1)}$ of the systematic error due to $R_{+\times}'(x)|_{x=0} \neq 0$ is given by the arclength of the portion of the great circle connecting the true source position and the position obtained by triangulation from $\delta^{(1)}_{13}$ and $\delta^{(1)}_{23}$ in Eqs. (\ref{eq:delta13}) and (\ref{eq:delta23}).
To give a representative example, $l^{(1)}$ is computed for the long monochromatic signal described above, for the HLV network with $\sigma^2_i = 1$, i.e. under the simplifying assumption of instruments with identical noises at all sites.
Only a very small fraction of the sky (about 0.7\%) has a negligible systematic error [$l^{(1)} (2\pi \times 40 \; {\rm Hz}/\omega) < 0.01$ rad]. 
It is therefore plain that position estimations will be grossly off target if the assumption that $R_{+\times}'(x)|_{x=0} = 0$ is wrongly made.

When $R_{+\times}'(x)|_{x=0} \neq 0$, more information about the waveforms is required to estimate the source position.
One possible approach is to use the {\tt CPF} algorithm, which assumes $R_{+\times}'(x)|_{x=0} = 0$, and then to correct the source position estimate using Eqs. (\ref{eq:delta13}) and (\ref{eq:delta23}).
This requires only a knowledge of the slope of $R_{+\times}$ with respect to the slope of $R_{++}$ at zero lag (i.e., $\omega_1^2/\omega_0$), which is closely related to the characteristic frequency of the waveforms.
A map from estimated position (with systematic error) to actual position must be constructed for every choice of $\omega_1^2/\omega_0$.
The remaining systematic error is given by the higher order terms not included in the correction.
For $\delta^{(2)}_{ij}$ the second order errors on the time delays, the second order equation derived from the linearization of Eq. (\ref{eq:normdeltaI}) is
\begin{equation}
\sum_{j=1, j\neq i}^N a_j [ F_{ij} \omega_1^2 \delta^{(2)}_{ij} - \omega_2^3 G_{ij} ( \delta^{(1)2}_{ij} + 2 \delta^{(1)}_{ij} \delta^{(2)}_{ij} ) ] = 0.
\end{equation}
This linear system of equations can be solved to obtain:
\begin{eqnarray}
\frac{-\delta^{(2)}_{13} D}{\delta^{(1)}_{13}} = a_1 \delta^{(1)}_{23} [ 2 a_2 \delta^{(1)}_{12} F_{12} G_{12} + a_3 \delta^{(1)}_{13} G_{13} ( F_{12} - 2\frac{\omega_2^3}{\omega_1^2} \delta^{(1)}_{12} G_{12} ) ] + a_3 \{ a_3 \delta^{(1)}_{12} \delta^{(1)}_{13} G_{13} [ F_{23} + 2\frac{\omega_2^3}{\omega_1^2} \delta^{(1)}_{23} G_{23} ] + \nonumber \\
 a_2 [ \delta^{(1)2}_{23} F_{12} G_{23} + \delta^{(1)}_{12} G_{12} ( \delta^{(1)}_{12} F_{23} + 2\frac{\omega_2^3}{\omega_1^2} \delta^{(1)}_{23} (\delta^{(1)}_{12} + \delta^{(1)}_{23} ) G_{23} ) ] \} 
\end{eqnarray}
and
\begin{eqnarray}
\frac{-\delta^{(2)}_{23} D}{\delta^{(1)}_{23}} = a_3 \delta^{(1)}_{23} G_{23} [ a_2 \delta^{(1)}_{13} ( F_{12} + 2\frac{\omega_2^3}{\omega_1^2} \delta^{(1)}_{12} G_{12} ) + a_3 \delta^{(1)}_{12} ( F_{13} + 2\frac{\omega_2^3}{\omega_1^2} \delta^{(1)}_{13} G_{13} ) ] + a_1 \{ 2 a_2 \delta^{(1)}_{12} \delta^{(1)}_{13} F_{12} G_{12} + \nonumber \\
a_3 [ \delta^{(1)2}_{13} F_{12} G_{13} + \delta^{(1)}_{12} G_{12} ( \delta^{(1)}_{12} F_{13} + 2\frac{\omega_2^3}{\omega_1^2} \delta^{(1)}_{13} (\delta^{(1)}_{12} - \delta^{(1)}_{13}) G_{13} ) ] \},
\end{eqnarray}
where
\begin{eqnarray}
D = a_3 \frac{\omega_1^2}{\omega_2^3} \{ a_1 \delta^{(1)}_{23} ( F_{12} - 2\frac{\omega_2^3}{\omega_1^2} \delta^{(1)}_{12} G_{12} ) ( F_{13} + 2\frac{\omega_2^3}{\omega_1^2} \delta^{(1)}_{13} G_{13} ) + \nonumber \\
(F_{23} + 2\frac{\omega_2^3}{\omega_1^2} \delta^{(1)}_{23} G_{23})[a_2 \delta^{(1)}_{13} (F_{12} + 2\frac{\omega_2^3}{\omega_1^2} \delta^{(1)}_{12} G_{12}) + a_3 \delta^{(1)}_{12}(F_{13} + 2\frac{\omega_2^3}{\omega_1^2} \delta^{(1)}_{13} G_{13}) ] \}.
\end{eqnarray}

\begin{figure}
\begin{center}
\ieps{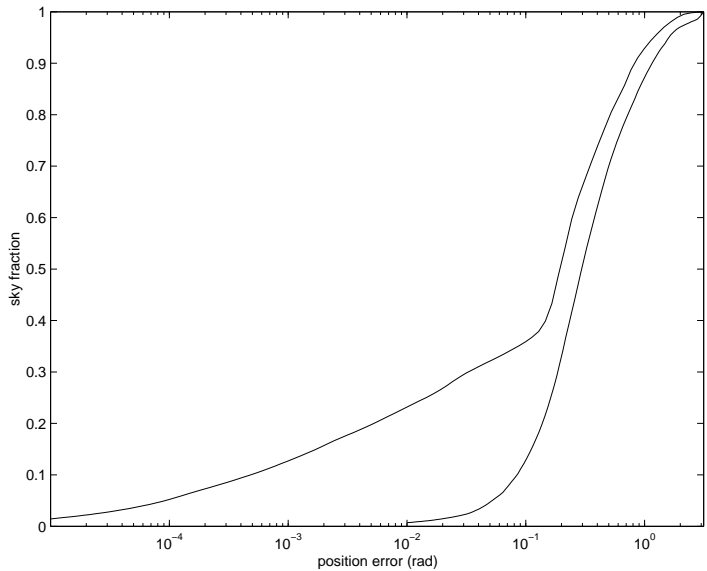}
\end{center}
\caption{The fraction of the sky with a systematic error smaller than the value of the position error plotted on the horizontal axis, for a long monochromatic signal of angular frequency $\omega = 80\pi$ rad/s, with two polarizations of the same amplitude, but at a phase offset of $\pi/2$. The rightmost curve is for $l^{(1)}$, the leftmost one for $l^{(2)}$.}
\label{fig:syserr}
\end{figure}

For $l^{(2)}$ the arclength of the portion of the great circle connecting the true source position and the position obtained by triangulation using $\delta^{(2)}_{ij}$, i.e., for $l^{(2)}$ the systematic error after correction using the first order expansion, and for the same example as above (with $\omega = 80\pi$ rad/s), one finds that 23.2\% of the sky has a negligible systematic error ($l^{(2)} < 0.01$ rad).
Figure \ref{fig:syserr} shows the fraction of the sky with a systematic error smaller than a certain value, for $l^{(1)}$ and $l^{(2)}$. 
It should be noted that for only about 50\% of the sky is it meaningful to use an expansion in the small parameters $\delta^{(1)}_{ij}$ and $\delta^{(2)}_{ij}$.
All in all, these numbers show that it is possible to use the {\tt CPF} algorithm for any signals, at the cost of limiting the sky coverage to $\sim 25\%$, and of requiring one additional piece of information about the signal, the value of the ratio $\omega_1^2/\omega_0$, which describes the behavior of the signal cross-correlation functions near zero lag.
Shorter signals, or signals with less overlap between their two polarizations, are likely to offer smaller systematic errors on position estimates, so that the correction suggested above becomes unnecessary.

\section{Numerical Simulations} \label{simulations}

All the numerical simulations were performed using the {\tt CPF} implementation for the LIGO Data Analysis System (LDAS), which uses the {\tt TFCLUSTERS} algorithm \cite{tfclusters} to measure the signal power in a given time-series.
The HLV network was analyzed, except that the noises of all three interferometers were assumed identical, for simplicity.
For every realization of the simulation procedure, a 10 s long segment of white noise of unit variance, sampled at 16384 Hz, was generated independently for all three interferometers. 
The use of white instead of colored noise is not an important loss of generality, since {\tt TFCLUSTERS} is quite robust with respect to the presence of correlations in the background noise.

Simultaneously, two 1/8 s long segments of unit variance white noise were generated, and filtered by a 6$^{\rm th}$ order elliptical bandpass filter with 3 dB cutoff frequencies at 125 Hz and 150 Hz, so that the amount of power outside this band was negligible after filtering.
These two segments were then truncated to their central 1/16 s long portion, and were used as the plus and cross polarizations of the gravitational wave signal incident on the network of detectors.
By construction, the average value of $\Lambda_{+/\times}$ was 1, and that of $\Lambda_{+\cdot\times}$ was zero.
In some simulations, the plus polarization waveform was multiplied by $\sqrt{2}$ and the cross polarization waveform was divided by $\sqrt{2}$, so that on average $\Lambda_{+/\times}$ was equal to 2.

All gravitational wave signals were injected from the position corresponding to the northern hemisphere normal of the HLV plane, so that they arrived in phase at the three interferometers.
The two polarizations were combined at each interferometer using the beam-pattern functions $F^+_i$ and $F^\times_i$, and were added directly to the background white noise.

A scale factor $A$ was multiplying the GW signal injected in the simulated data for all interferometers.
If the frequency band occupied by the signal, its duration, and its arrival time had all been known exactly, it would have been possible to filter the data optimally in that band.
In that case, the signal-to-noise ratio $\rho_{\rm opt}$ for a certain scale factor $A$ would have been given by 
\begin{equation}
\rho_{\rm opt}(A) = A \sqrt{\sum_{i=1}^3 (F^{+ 2}_i + F^{\times 2}_i)}. \label{eq:snropt}
\end{equation}
Of course, it cannot be assumed that this information is available, but I will nevertheless use Eq.(\ref{eq:snropt}) as a measure of the strength of the injected signals, instead of using $A$ which is less intuitive.
Numerically, $\rho_{\rm opt}(A) = 1.34 A$ for a source at the northern hemisphere normal of the HLV plane.

The simulations were performed using a realistic hierarchical implementation of {\tt CPF}. 
The simulated data from the three interferometers were first processed separately by {\tt TFCLUSTERS} to produce three lists of events.
In terms of the notation developed in \cite{tfclusters}, the settings of {\tt TFCLUSTERS} were $\alpha = 0$, $\sigma=5$, and $\vc{\delta} = [0,0,0,0,0,0,2,3,4,4]$.
Only the frequencies below 1024 Hz were considered, and the time resolution of the time-frequency decomposition was $T=1/8$ s.
The number of events in this first stage was controlled by the black pixel probability threshold, $p_0$ (note that larger values of $p_0$ give {\em larger} false alarm rates).

The events produced by {\tt TFCLUSTERS} are in the form of rectangles in the time-frequency plane, with information about the power present in each pixel of these rectangles.
For a given cluster identified by {\tt TFCLUSTERS}, the rectangle is defined as the smallest rectangle containing all the pixels of the cluster.
A triple-coincidence condition was therefore applied as in \cite{THESE}: to be coincident, the time-frequency rectangles corresponding to the events from all three interferometers had to be overlapping.
This coincidence condition selects events that are close in time and in frequency, and can be understood as a standard time and frequency coincidence gate with varying windows that are fixed by the events under consideration.

All the coincident events that were present in a given 10 s segment were then considered in turn.
Their start time, duration, central frequency and bandwidth were estimated from the smallest rectangle in the time-frequency plane that could contain the union of the rectangles from the individual events.
The {\tt CPF} algorithm was then ran on the data, once for each coincident event.
The implementation employed used {\tt TFCLUSTERS} to process the synthetic data, and the measure of the power was the sum of the power in all the pixels identified by {\tt TFCLUSTERS} that were inside the time-frequency rectangle identified from the triple coincidence.
The parameters for {\tt TFCLUSTERS} were the same as those mentioned above, except for the black pixel probability, which was set to a value $p_1$.
The threshold defined by $p_1$ was chosen so that only loud enough signals were detected by {\tt TFCLUSTERS}, and their estimated power was linearly related to their actual power.

The power measured by {\tt TFCLUSTERS} was maximized over the source position (two angles) and over the parameter $\Lambda_{+\cdot\times}$.
It was assumed that the value of $\Lambda_{+/\times}$ was known beforehand, in order to keep the size of the parameter space small enough for simulations.
In a first time, the sky was covered by picking 100 points uniformly distributed in the range $[0,2\pi[$ for the right ascension, and 100 points in $[-1,1[$, uniform in the sine of the declination angle.
In addition, 10 points were used to cover the range $[-1,1[$ uniformly for the parameter $\Lambda_{+\cdot\times}$.
Consequently, ${\tt TFCLUSTERS}$ was ran $10^5$ times on every 10 s long simulation. 
Including the overhead from the LDAS system, this part of the search ran in $\sim 225$ s on 31, 2 GHz Pentium IV computers, with 512 Mb of RAM.

If none of the triple coincidence events registered above the $p_1$ threshold of {\tt TFCLUSTERS} when analyzed by {\tt CPF}, a non-detection was reported and the analysis was stopped.
Otherwise, a detection was announced, and {\tt CPF} produced a scan of the parameter space for every triple coincidence event above threshold.
The triple coincidence event with the largest maximum power was then selected as a possible GW candidate, and was analyzed in more details.
The point in parameter space where the power was maximum defined the parameters for a second run of {\tt CPF}, used to obtain a refined position estimate not limited by the coarseness of the grid covering the parameter space.
The value of $\Lambda_{+\cdot\times}$ was fixed to the value estimated in the first run, and a square search window of size 0.2 rad in right ascension and in declination, with 50 steps in both angles, was centered on the value of the position obtained in the first run.
The position with the maximum power in this second run was taken as the final estimate of the source position.

\subsection{Detection}
For the sole purpose of detecting the presence of a signal in the data, only the first run of {\tt CPF} was required.
Data were ran through {\tt TFCLUSTERS} separately, the events were combined in the triple coincidence gate, and the data were fetch through {\tt CPF} for the parameters defined by each triple coincidence event.
If at least one of the triple coincidence event lead to a detection by {\tt CPF}, the 10 s long time interval under scrutiny was assumed to contain a signal.
By design, this hierarchical scheme required a fairly permissive threshold in the first stage where {\tt TFCLUSTERS} was ran independently on every interferometer, so that a given signal was very likely to make it to the second stage where {\tt CPF} was ran.
The limit on this threshold was determined by the availability of computational resources, and by the confusion that resulted from the proliferation of events at low threshold.
Most of the rejection of accidental coincidences occurred at the second stage, where {\tt CPF} was operated with a reasonably strict threshold.

A numerical experiment was performed by running this simulation a large number of times, with and without signal injection.
When injected, the signal had $\rho_{\rm opt} = 13.4$ and $\Lambda_{+/\times} = 1$.
The thresholds were chosen to be $p_0 = 0.14$ and $p_1 = 0.012$.
In the first stage (triple coincidence), the probability to detect a signal was $P_D = 0.92 \pm 0.01$, and the probability of a false alarm when no signal was present was $P_F = 0.62 \pm 0.02$.
In the second stage ({\tt CPF}), it was measured that $(P_D, P_F) = (0.86 \pm 0.01, 0.10 \pm 0.01)$.
Overall, it was measured with both staged combined that $(P_D, P_F) = (0.79 \pm 0.02, 0.064 \pm 0.009)$.
Table \ref{tab:detprob} gives the detailed results from the simulations.
The errors quoted here come from 68.3\% confidence intervals (``$1\sigma$'') for a Bernoulli process, built using the prescription of \cite{FC}.
The 6.4\% probability of false alarm is larger than typical values for GW searches, as it would give a false alarm rate around 7 mHz.
It was chosen, however, in order to provide enough detections for small errors on the measured probabilities.
In a more realistic setting, $p_0$ would be similar to the value used here, while $p_1$ might be smaller by one or two orders of magnitude.
It should be noted that the threshold settings were found by trial and error; there is an infinity of points along a curve in the $p_0,p_1$-plane that give a 6.4\% probability of false alarm, and a choice different than the one above may give a larger probability of detection.
\begin{table}
\begin{center}
\begin{tabular}{|c|c|c|} \hline
 & no signal & injected signal \\ 
 & injected & with $\rho_{\rm opt} = 13.4$ \\ \hline 
total number & 981 & 730 \\  \hline
detected by & 611 & 669 \\ 
triple coincidence & [0.606,0.640] & [0.904,0.927] \\ \hline
undetected by triple & 370 & 61 \\
coincidence & [0.361,0.394] & [0.0726,0.0962] \\ \hline
detected by {\tt CPF} & 62 & 578 \\
 & [0.0549,0.0731] & [0.775,0.808] \\ \hline
undetected by {\tt CPF} & 549 & 91 \\
 & [0.542,0.576] & [0.112,0.139] \\ \hline
\end{tabular}
\caption{Details of the simulations to measure the efficacy of the hierarchical implementation of the {\tt CPF}. Numbers in brackets show the 68.3\% (``$1\sigma$'') confidence interval for the fraction of the number of trials to the total number of trials.}
\label{tab:detprob}
\end{center}
\end{table}

Nevertheless, it seems that the choice I made for the $p_0$ and $p_1$ thresholds is sufficient to show the superiority of the coherent approach over the incoherent one for detection.
I repeated the experiment above, but using only the first stage to detect events. 
A detection was announced when at least one triple coincidence was observed between the outputs of the three {\tt TFCLUSTERS} runs on the interferometers' data.
In that case, a threshold of $p_0 = 0.11$ gave $(P_D, P_F) = (0.70 \pm 0.02, 0.07 \pm 0.01)$, in an experiment with 742 trials for the measurement of $P_D$ (519 detections), and 742 trials for $P_F$ (48 detections).
At a similar false alarm probability, the probability of detection is significantly smaller in the incoherent case than in the coherent case.
For the particular signal and false alarm probability under consideration, the signal-to-noise ratio would have to be increased to $\rho_{\rm opt} = 16.8$ in order for the incoherent approach to be as efficient as the coherent one.
For a homogeneous distribution of sources in space, this corresponds to a factor of $\sim 2$ improvement in detection rate, assuming no significant degradation of the {\tt CPF} algorithm performances with respect to the incoherent algorithm as the position of the source is varied away from the northern hemisphere normal to the HLV plane.

The performances of the {\tt CPF} search were mostly limited by the quality of the estimation of the time-frequency rectangle containing the burst, in the first stage of the analysis (the triple coincidence).
With the signal injection for $\rho_{\rm opt} = 13.4$, the first stage gave an estimated rectangle that overlapped with the one containing the signal in 96\% of the 578 cases where the signal was discovered, but only in 3\% of the 61 missed detections was this the case.
Without signal injection, only 1\% of the 611 triple coincidences had time-frequency rectangles overlapping with the signal rectangle.
These numbers show that if an oracle were available to provide the rectangle containing the signal without error every time the search is ran, a probability of detection $\agt 97\%$ would be possible for a probability of false alarm $\alt 1\%$.
Stated differently, {\tt CPF} is extremely efficient at detecting a burst when it receives the right parameters describing that burst; the triple coincidence incoherent search provides many candidates; when one such candidate corresponds to the signal, {\tt CPF} picks it out of the others very efficiently.
A better approximation to this oracle than the one used here might be to tile the time-frequency plane with a variety of rectangles, and to run {\tt CPF} on each rectangle.

Suppose that it is known that the signal has a bandwidth of 25 Hz.
One can cover the time-frequency plane with non-overlapping rectangles of duration 1/16 s, and bandwidth of 25 Hz, so that the 10 s long, 1024 Hz bandwidth data segment in my simulations is covered by $\sim 6500$ tiles.
The false alarm probability for each run of {\tt CPF} must be reduced to $\sim 10^{-5}$ so that the global search has $P_F \sim 7\%$.
The efficiency of the search would then approximately be given by the probability of detection of {\tt CPF} with an oracle, for a threshold giving $P_F = 10^{-5}$.
I have not measured this probability of detection, but it is plausible that it is larger than the 79\% efficiency measured for the hierarchical implementation used in the simulations.
However, running this search in real time would be prohibitively expensive, as it would require $\sim 10^4$ teraflops of computational power.
In the present hierarchical implementation, the first stage took $\sim 30$ gigaflops to run in real time, and the second stage, $\sim 800 (P_F/0.62)$ gigaflops, where $P_F$ is the false alarm probability in the first stage.
These numbers should be taken as upper bounds on the required computational power, because the codes were not optimized to minimize overhead, to make an optimal usage of the parallel resources, or to do an optimal scan of the parameter space.
Optimized codes should be able to run at least $\sim 5$ times faster.

\subsection{Position estimation}
In order to get a precise idea of the magnitude of the position estimation errors, the analysis scheme described above was simplified by removing the first incoherent step involving the three different instances of {\tt TFCLUSTERS} running at each site.
Instead, the {\tt CPF} algorithm was instructed to compute the power according to the output of {\tt TFCLUSTERS} in a rectangle of duration 1/8 s located at the right position in the time series, with a lower frequency of 50 Hz and an upper frequency of 150 Hz.
The black pixel probability was set to $p_1 = 5\times 10^{-3}$, so that the number of clusters unrelated to the signal and produced only by the noise was small.

\begin{figure}
\begin{center}
\ieps{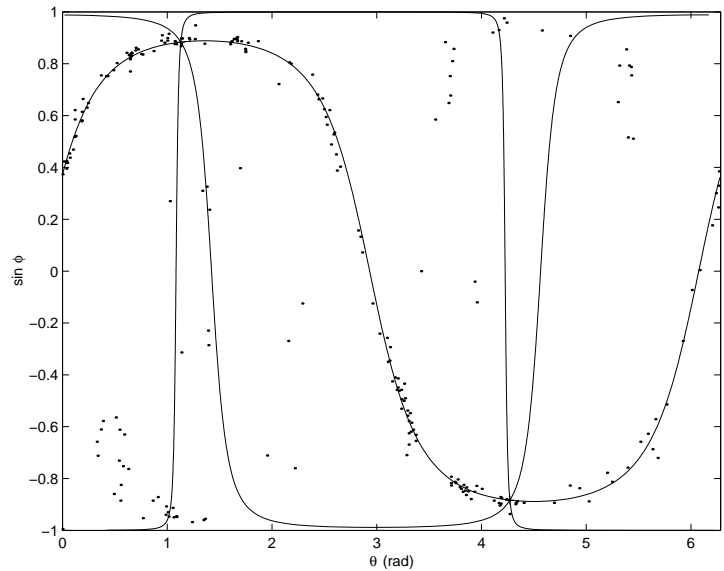}
\end{center}
\caption{Scatter plot of the estimated position of a source injected along the northern hemisphere normal of the HLV plane, from 240 realizations of the simulation with $\Lambda_{+/\times} = 1$ and $\rho_{\rm opt} = 13.4$. Horizontal axis is right ascension, vertical axis is the sine of the declination. The curves represent loci of equal time delay for the three independent interferometer pairs. The point where they intersect in the upper left corner of the figure is where the signal was injected.}
\label{fig:posU10}
\end{figure}

Figure \ref{fig:posU10} presents a scatter plot of the position estimates obtained for 240 realizations of the simulation, when $\Lambda_{+/\times} = 1$.
The estimates tend to cluster along the curve corresponding to the locus of positions having equal delays at the Hanford and Livingston interferometers.
This is a direct consequence of the good alignment between these two interferometers, and the relatively poor alignment of the Virgo detector with them.
Estimates tend to fall on that curve, but also to cluster at different places along it, where the signals at Hanford and at Livingston are delayed by an integer number of the characteristic periods of the signal with respect to the signal at Virgo.

\begin{figure}
\begin{center}
\ieps{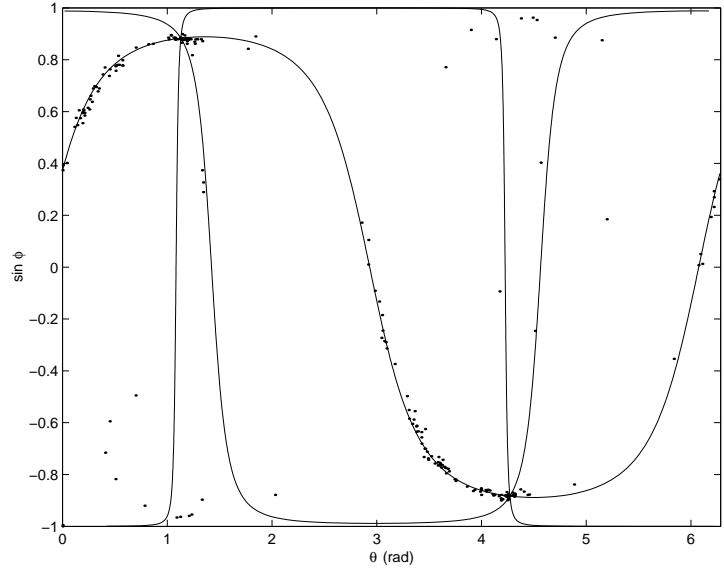}
\end{center}
\caption{Same as Figure \ref{fig:posU10}, but with $\Lambda_{+/\times} = 2$.}
\label{fig:posU210}
\end{figure}

Figure \ref{fig:posU210} shows a similar plot as Fig.\ref{fig:posU10}, except that $\Lambda_{+/\times} = 2$.
This corresponds to a GW signal which has more structure in its polarizations than the one for the case $\Lambda_{+/\times} = 1$, i.e. which is closer to a linearly polarized signal.
Since linearly polarized signals are the easiest ones to analyze with a network of interferometers, it is expected that the position estimates will be better.
This is indeed was is observed in Fig.\ref{fig:posU210}; the estimates still hug the Hanford-Livingston equal-delay curve, but now present less scatter around the points where the signals are in phase with the Virgo signal along that curve.

\begin{figure}
\begin{center}
\ieps{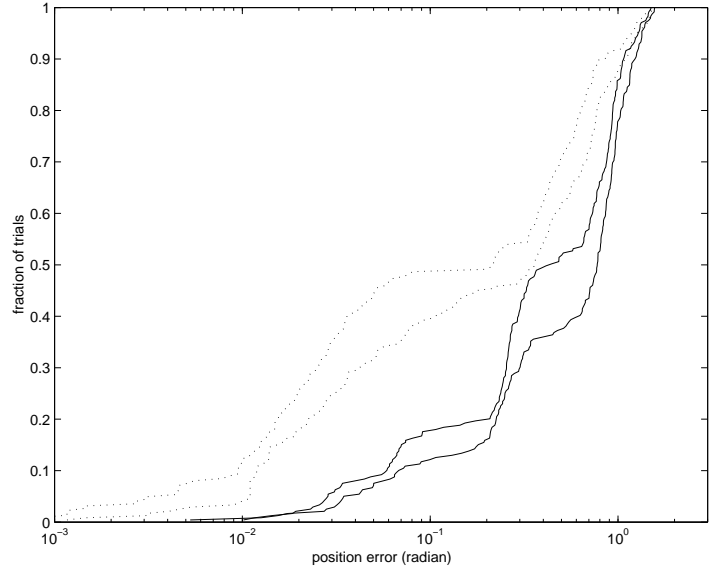}
\end{center}
\caption{The fraction of all simulations that gave a position error smaller than the value plotted on the horizontal axis. The two continuous lines correspond to $\Lambda_{+/\times} = 1$, and the two dotted lines to $\Lambda_{+/\times} = 2$. In both cases, the rightmost curve is for $\rho_{\rm opt} = 13.4$, and the leftmost one is for $\rho_{\rm opt} = 35.6$. Each curve is built from 240 realizations of the simulation. The error on the curves is estimated to be $\sim 5\%$ of the fraction of trials.} 
\label{fig:posErr}
\end{figure}

The position error can be quantified as in Fig.\ref{fig:syserr}: it is taken to be the length of the shortest portion of the great circle joining the estimated position and the true source position or its mirror image with respect to the HLV plane.
The error can be defined with respect to the mirror image because there is a natural ambivalence in the estimation of the position when only three detectors are used.
Figure \ref{fig:posErr} shows the cumulative distribution of the position error from the simulations, for $\Lambda_{+/\times} = 1$ and for $\Lambda_{+/\times} = 2$, and for two different values of the signal-to-noise ratio of the injected signals.
As expected, signals with larger values of the signal-to-noise ratio or of $\Lambda_{+/\times}$ lead to smaller position errors.
The curves in Fig.\ref{fig:posErr} present a number of ``steps'', which are produced by the clustering along the Hanford-Livingston equal-delay curve at positions in phase with Virgo.

Roughly 50\% of the trials lead to unusable position estimates (errors $\agt$ 10 degrees) when $\Lambda_{+/\times} = 2$, while this number reaches $\sim 80\%$ when $\Lambda_{+/\times} = 1$.
However, with $\Lambda_{+/\times} = 2$, approximately 25\% of the trials have errors smaller than one degree.
Moreover, at least in the regime of signal-to-noise ratios under consideration here, the scaling of the position error with the signal-to-noise ratio is rather weak.

\section{Conclusion}
A method was presented for the optimal generalization of the power detectors developed for single interferometers so that they can process coherently data from a network of interferometers.
The coherent method, as compared to an incoherent approach where event lists independently generated at all interferometers are searched for coincidences, offers the advantage of better detection efficiency, and the possibility to accurately estimate the position of the source.
A few systematic effects affect the performances of the {\tt CPF} algorithm for position estimation, including: cross terms between the plus and cross polarization waveforms of the GW signal, lack of differences between the characteristics of the two polarization waveforms, and misalignment of the interferometers of the network.

The three effects are related, and reflect the obvious fact that GW signals incident on a network of misaligned detectors will only show entangled versions of their two polarization waveforms, different in each interferometer.
If one of the two polarizations is significantly stronger than the other (i.e., the GW signal has a stronger degree of linear polarization), the problem is drastically simplified.
Similarly, aligned interferometers are much less sensitive to this problem.
If the two polarization waveforms are fairly coherent with each other and of similar amplitudes, the cross terms between the two polarizations in different interferometers may show significant maxima when the time shifts imposed on the different data streams do not correspond to the differences in time-of-arrival of the GW signals at each interferometer.
These maxima cannot be distinguished from the maximum resulting from the product of the waveforms of the same polarization in different interferometers, and systematic position errors may result.
It should be noted, however, that the same effect leads to similar position errors when the differences in arrival time of the GW signal in different interferometers are estimated by maximizing the cross-correlation between pairs of interferometers, and these time differences are used to triangulate the source position.

The study of the {\tt CPF} algorithm presented here was based on a three interferometer network.
However, the structure of the algorithm allows for any number of interferometers.
More complex networks will most likely reduce the effects of the systematic errors, by increasing the number of linear combinations of the two polarization waveforms that are being sampled, or, equivalently, by increasing the area of the sky where at least three interferometers are measuring similar combinations of the two polarizations.
Numerical simulations on the network formed by the LIGO Hanford, LIGO Livingston, and Virgo interferometers (the HLV network) with a short random signal showed that the {\tt CPF} algorithm could be used to accurately measure the position of the source a significant fraction ($\sim 25\%$) of the time, for reasonably strong sources located at the normal of the HLV plane. 
In about $\sim 90\%$ of the trials, the {\tt CPF} algorithm correctly placed the source position at a point where the signals at the Hanford detector and at the Livingston detector were in phase.
However, due to its misalignment with respect to the LIGO detectors, the information provided by the Virgo detector was often insufficient to pinpoint correctly the position of the source.
The average position error was a rather weak function of the strength of the signal, but a stronger function of the amount of difference in the structure of the two polarization waveforms.

It was also shown that the {\tt CPF} algorithm offers better detection efficiencies than its incoherent equivalent, both for directed and for all-sky blind searches, and independently of the systematic errors affecting the position estimations.
In the former case, improvements in the detection signal-to-noise ratio of 40\% or better are expected, excepted for a few values of the source parameters.
In the latter case, a 25\% improvement in signal-to-noise ratio was measured for typical source parameters.
This improvement for the all-sky blind search comes at the cost of increasing the computational power required to perform the data analysis in real time by a factor of $\sim 5-30$.
This may be significant, especially since lengthy simulations probably have to be completed in order to estimate the false alarm and detection statistics of a real search.
However, the signals studied in the numerical simulations, which could be viewed as very rough approximations to the signals that could be emitted in the collapse of the core of a star in supernova explosions, would have been detected by the {\tt CPF} algorithm at a rate $\sim 2$ times larger than the rate for its incoherent equivalent, assuming a homogeneous spatial distribution of the sources, and that the performances of the {\tt CPF} algorithm for a source injected at the normal of the HLV plane are characteristic of its performances at other injection positions.

Finally, an important advantage of the design of the {\tt CPF} algorithm is that it inherits the robustness, efficacy, and computational efficiency of the single interferometer power detectors.
In other words, the {\tt CPF} algorithm should not be more sensitive to non-Gaussian or non-stationary noises than incoherent analyzes, it should be searching more efficiently for GW signals sharing similar morphologies, and, while more computationally intensive, its speed should scales the same way with improvements in the implementation of the single interferometer power detectors, or with computing hardware ameliorations.

\acknowledgments 
This work was supported by the National Science Foundation under cooperative agreement PHY-9210038 and the award PHY-9801158.
This document has been assigned LIGO Laboratory document number LIGO-P030007-01-D.

\end{document}